\documentclass[twocolumn,aps, superscriptaddress]{revtex4-1}

\usepackage{graphicx}

\usepackage{amssymb}
\usepackage{amsmath}
\usepackage[latin1]{inputenc}
\usepackage{bm}

\newcommand{\xib}{\boldsymbol{\xi}}

\begin{document}

\title{Gate driven adiabatic quantum pumping in graphene}

\author{E. Prada}
\affiliation{Instituto de Ciencia de Materiales de Madrid (CSIC),
Cantoblanco, 28049 Madrid, Spain}
\author{P. San-Jose}
\affiliation{Instituto de Estructura de la Materia (CSIC), Serrano 123, 28006
Madrid, Spain}
\author{H. Schomerus} \affiliation{Department of Physics,
Lancaster University, Lancaster,  LA1 4YB, United Kingdom}

\date{\today}

\begin{abstract}
We propose a new type of quantum pump made out of graphene, adiabatically
driven by oscillating voltages applied to two back gates. From a practical
point of view, graphene-based quantum pumps present advantages as compared to
normal pumps, like enhanced robustness against thermal effects and a wider
adiabatic range in driving frequency. From a fundamental point of view, apart
from conventional pumping through propagating modes, graphene pumps can tap
into evanescent modes, which penetrate deeply into the device as a
consequence of chirality. At the Dirac point the evanescent modes
dominate pumping and give rise to a universal response under weak driving for
short and wide pumps, even though the charge per unit cycle in not quantized.
\end{abstract}

\maketitle

\section{Introduction}
Externally driven quantum devices provide an attractive setting to transfer
charges between electronic reservoirs \cite{Buttiker:ZPB94, Brouwer:PRB98,
Makhlin:PRL01, Aharonov:PRL87}. Conventional realizations of such quantum
pumps  are based on patterned semiconductor heterostructures, and commonly
utilize the adiabatic or non-adiabatic loading and unloading of electrons
into well-defined individual quantum states, where energy gaps are provided
by Coulomb interactions or by size quantization
\cite{Kouwenhoven:PRL91,Pothier:EL92,Switkes:S99,Watson:PRL03,Blumenthal:NP07,Kaestner:PRB08,Wright:PRB09,Pekola:NP07}.

Graphene, consisting of an atomically thin sheet of carbon atoms
\cite{Novoselov:S04}, provides a novel platform for two-dimensional electronic
transport which combines a regime of nominally vanishing charge carrier
density (at the Dirac point) with radically different confinement properties.
These properties are intimately linked to the chiral nature of the charge
carriers, which suppresses backscattering at interfaces and results in
the so-called Klein paradox by which charge carriers are difficult to confine
\cite{Neto:RMP09, Katsnelson:NP06, Beenakker:RMP08}. Here we show that
chirality and vanishing carrier density conspire to yield a universal
dimensionless pumping efficiency at the Dirac point, which remarkably
corresponds to a non-quantized charge per unit cycle \cite{Prada:PRB09}. The
pumping then is mediated by evanescent modes, which previously have been
studied in the context of stationary transport \cite{Katsnelson:EPJB06,
Tworzydlo:PRL06,Schomerus:PRB07}. We also analyze the behavior of graphene pumps away from
the Dirac point \cite{Prada:PRB09,Zhu:APL09,Tiwari:PA}, which is dominated by
propagating modes, and compare it to that of normal quantum pumps. 
In the second part of this paper we turn to issues of practical implementations.
We find that graphene's spectral 
properties enhance the response to external driving by gates, 
increase the robustness against thermal
effects, and result in a wider adiabatic range in driving frequency. Graphene pumps therefore
offer a number of potential advantages which may prove useful for practical applications.

This work is organized as follows.
Section \ref{sec:2}  introduces a model of a quantum pump driven by two
electrostatic gates, which we analyze in the scattering approach to adiabatic
quantum pumping. Results for normal and graphene-based pumps are presented in
Sec.\ \ref{sec:3}, where we formulate them  in terms of the onsite potential
driving. In Sec.\ \ref{sec:4} we apply these results to the experimentally
relevant case of driving via gate voltages. Section \ref{sec:5} contains
estimates of the maximal pumping performance of realistic graphene devices
and their normal counterparts. We conclude in Sec.\ \ref{sec:6} with a
summary of the practical advantages of graphene-based quantum pumps.

\begin{figure}
\includegraphics[width=\linewidth]{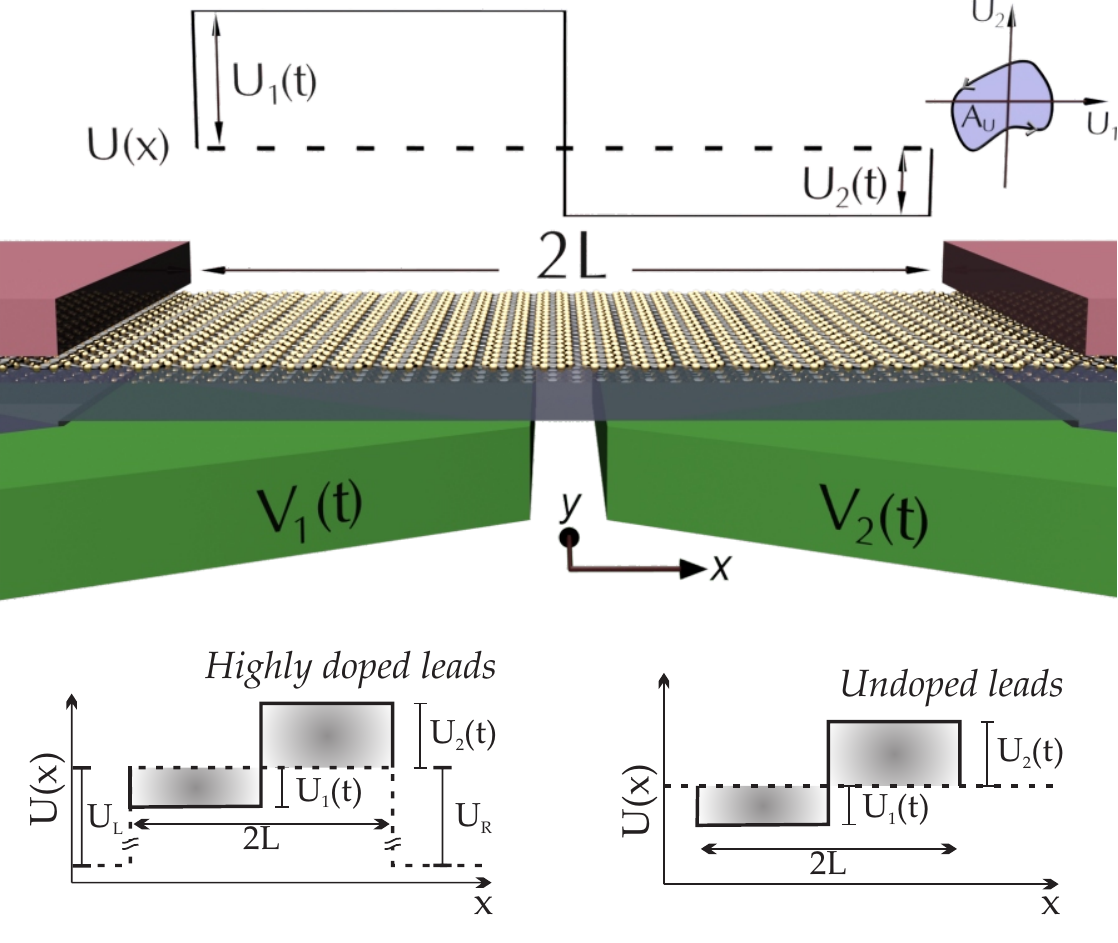}
\caption{\label{fig:system} (Color online) Top panel: Graphene quantum pump where two gates at voltages $V_1(t)$ and
$V_2(t)$ induce a periodic time-dependent onsite energies $U_1(t)$ and
$U_2(t)$ in the two halves of a graphene flake.
We investigate the resulting charge transport between two contact electrodes, separated by
a distance $2L$. Bottom panels: Instantaneous onsite potential at a point during the pumping cycle (solid thick line), including the potential in the contact region for highly doped contacts (left, $U_{L,R}\rightarrow -\infty$) and undoped contacts (right). The dashed line shows the potential at the working point (in the absence of driving).}
\end{figure}

\section{\label{sec:2} Model and framework}

In order to pump an average current between two reservoirs that are kept at
the same bias, it is necessary to vary the scattering properties of the pump
region (the system between the reservoirs) periodically over time. This is
achieved by driving the system with some sort of external forces that modify
the parameters of the system, hence the name `parametric pumping'.
The driving forces can be induced by any external parameter (like a
magnetic field, the transparency of the contact between the pump and the
reservoir, the potential energy, etc.), as long as their variation modifies the
scattering matrix of the system.
When the
temporal variation of these parameters is sufficiently slow (as compared to
the dwell time of the carriers within the pump region), the pump operates in
the adiabatic regime. To each traversing electron, the system then is
approximately static, and the
scattering amplitudes become insensitive to the explicit time dependence of the
driving.   In this case, the pumped charge increases
proportionally to the frequency of the external driving, and at least two
parameters of the system must be driven out of phase. The direction of the
pumped current depends on the specific way the scattering phases depend on
the varying parameters, and  changes sign when the driving cycle is
reversed.

In this work we consider a new type of adiabatic quantum pump, made out of
graphene, and compare it to the equivalent normal pump. We propose to
externally drive the pump by the capacitative coupling to two adjacent gates.
Such devices have already been
experimentally investigated for stationary transport, both for graphene as well as for `normal' conductors (such
as semiconducting quantum wells). Oscillating gate voltages $V_1(t)$ and
$V_2(t)$ induce an oscillating shift $U_1(t)$ and $U_2(t)$ in the onsite
potential of each of the two adjacent regions coupled to the gates.
These two oscillating potential barriers have a width $W$ and a length $L$,
with a total pumping region of length $2L$ \footnote{Note that Ref. \onlinecite{Prada:PRB09}, which discusses this same setup, has a typo in Fig. 1, where the length of the system is labeled as $L$, instead of the correct $2L$.}. A sketch of the described setup,
attached to two metallic contacts (left and right reservoirs), can be seen in
the upper panel of Fig.~\ref{fig:system}.


We characterize the unique features of quantum pumping in graphene by comparison of
four different variations of the setup above:
graphene pumps with either heavily doped contacts (bottom left panel of Fig.
\ref{fig:system}) or undoped contacts (bottom right panel)  are contrasted
with pumps where the graphene flake is replaced by a normal system accommodating an ordinary two-dimensional electron gas.
In the absence of driving, i.e., at the
`working point' of the pump (dashed line in the bottom panels of Fig.
\ref{fig:system}),  systems with undoped contacts have a uniform carrier
density, equal in both the leads and in the central pumping region. In
contrast, the charge carrier density in heavily doped leads is very large;
this can be realized by metallic electrodes. The charge carrier density in the system can be uniformly shifted by
another back (or top) gate covering both the pumping region and the leads (not
shown in Fig.\ \ref{fig:system}), which allows to tune the working point of the
device. For heavily doped contacts, the back-gate-induced change of the
charge carrier density in the leads can be neglected.
The different types of contacts allow us to
isolate the influence of evanescent modes, which are only induced by
the highly doped contacts.

%

We follow the scattering
approach to quantum pumping \cite{Brouwer:PRB98, Makhlin:PRL01}, which relates the adiabatic  transfer of non-interacting charge carriers
to the parametric evolution of the instantaneous
scattering matrix $S(t)$. For
two independent pumping parameters $\xib=\{\xi_1, \xi_2\}$ (the minimal requirement for adiabatic pumping) and single
channel reservoirs, the charge transferred across the
scattering region from the left reservoir reduces to an integral over the area $A$ enclosed by the
driving path in two-dimensional parameter space,
\begin{subequations}\label{Q}
\begin{eqnarray}
Q&=&\int_A d\xi_1 d\xi_2\,\, \partial^2_{\xi}Q(\xib),\\
\partial^2_{\xi}Q
&\equiv&\frac{e}{\pi}\sum_{j=L,R}\mathrm{Im}\frac{\partial S^{*}_{L,j}}{\partial \xi_1}\frac{\partial S_{L,j}}{\partial\xi_2},
\label{Qb}
\end{eqnarray}
\end{subequations}
where $e$ is the electron charge, and $L,R$ indicate left and right contacts. The scattering matrix can be expressed in terms of the transmission probability through the system $T$ and the scattering phase $\phi=\alpha-\beta$ given by
\[
S=
\left(\begin{array}{cc}
r & t'\\
t& r'
\end{array}\right)
= e^{i\gamma}\left(\begin{array}{cc}
\sqrt{1-T}e^{i\alpha} & -\sqrt{T}e^{i\beta}\\
\sqrt{T}e^{-i\beta} & \sqrt{1-T}e^{-i\alpha}
\end{array}\right),
\]
where $r$ ($r'$) and $t$ ($t'$) are the $\xib$-dependent reflection and transmission
amplitudes for electrons arriving from the left (right) reservoir.
For wide quantum pumps depicted in Fig. \ref{fig:system}, the number of channels is large. However, because of the
quasi one-dimensional design where the onsite potential $U$ is independent of the transverse coordinate $y$,
the different channels remain decoupled.
Indexing the channels by a quantum number $n$,
the total pumped charge is therefore given by a sum $Q=\sum_{n}Q_{n}$.

Into the  longitudinal $x$-direction, we model the potential
profile by two abrupt steps of equal length
$L$ and assume that the two driving parameters $\xib=\{U_1(t),U_2(t)\}$ have
zero average, with maximum amplitudes $\delta U_1$ and $\delta U_2$,
respectively. Undoped contacts have the same Fermi
momentum $k_F$ as the pumping region, while heavily doped contacts have a
much larger Fermi momentum, which can effectively be taken as infinite
\footnote{Consequently, all our results do not feature an
additional Fermi momentum for the leads.}.

For this set-up, the  transmission probabilities $T_{n}$  and phases   $\phi_{n}$
can be computed by simple wave matching, where graphene regions are described by
the Dirac equation (with Dirac velocity $v_F$) \cite{Neto:RMP09}, and normal regions by the Schr{\"o}dinger equation
for quasiparticles of effective mass $m^*$. The results depend on the characteristic energy scale for the longitudinal confinement
of the carriers, given as $E_L^G\equiv\hbar v_F/L$ in the graphene case, and
$E_L^N\equiv\hbar^2/(2m^*L^2)$  for the normal conductor. Depending on the
amplitude $\delta u_i$ of the dimensionless driving energies $u_i\equiv
U_i/E_L$, the driving can be characterized as weak ($\delta u_i\ll 1$) or strong ($\delta u_i\gg 1$). For weak driving, the charge
\begin{equation}\label{weakdriving}
  Q_{n}\approx\partial_{u}^2 Q_{n}(0)\int d u_1 d u_2=A_{u} \partial_{u}^2 Q_{n}(0)
\end{equation}
 pumped in
channel $n$ becomes proportional to the small area $A_{u}\sim \delta u_1\delta u_2$ enclosed
in parameter space by the driving cycle, wherein $\partial_{u}^2
Q_{n}(u_1,u_2)$ can be approximated by a constant.
For strong driving the integral in Eq.\ (\ref{Q}) has to be
performed numerically.

\section{\label{sec:3} Results and discussion}

Let us first consider the case of graphene with heavily doped contacts. There are four regions in the scattering problem: the left contact ($x<-L$), the region of the first barrier with onsite potential $u_1$ ($-L<x<0$), the region of the second barrier with onsite potential $u_2$ ($0<x<L$), and the right contact ($x>L$). The transport problem across the pump can be solved by matching the propagating and evanescent modes to the left and right of the interfaces separating different regions, which
we obtain from the equation $H\Psi = (E-U_i)\Psi$. Here $H$ is the Dirac Hamiltonian $H=v_F \vec{\sigma} \cdot \vec{p} $, where  $\vec{p}$ is the momentum operator relative to the Dirac point and $\sigma _i$ are the Pauli matrices. The Dirac Hamiltonian acts on a two-component spinor, $\Psi=(\phi _A,\phi_ B)^T$, representing  the
amplitude of the wavefunction of energy $E$ on the two inequivalent triangular sublattices of graphene, labeled  $A$ and $B$.
The scattering at each interface conserves energy $E\equiv E_L \epsilon$ and the component $p_y\equiv\hbar q$ of the momentum parallel to the interface, where the latter plays the role of channel index $n$ (note, however, that due to spin and valleys, each value of $q$ has degeneracy $g=4$). In the left and right contacts, where $U_{L,R} \rightarrow -\infty$, a mode propagating towards the right (like the incident and transmitted ones) is proportional to  $(1,1)^T$, while a propagating mode moving towards the left (the reflected one) is proportional to $(1,-1)^T$. In the pump region, the scattering wave function with transverse wave vector $q$ is given by
\begin{eqnarray}
\Psi^{\pm}=e^{i(qy\pm k_x x)}
\left(\begin{array}{cc}
\lambda\\
z^{\pm}
\end{array}\right),
\end{eqnarray}
where $\lambda=\pm$ is positive for electron-like and negative for hole-like quasiparticles, $z^{\pm}\equiv(\pm k_x+iq)/\sqrt{k_x^2+q^2}$, and $\pm k_x=\pm\sqrt{[(\epsilon-u_i)/L]^2-q^2}$ is the electron's longitudinal momentum along the
transport direction ($+$ for a quasiparticle moving towards the right, $-$ for a quasiparticle moving towards the left).

To facilitate the calculation, we first solve the scattering problem of a single potential barrier, $u_1$ for example, and then find the result for the double-barrier problem by composition of scattering matrices. The wave matching condition of continuity for a single barrier at $x=0$ is
\begin{eqnarray}
\left(\begin{array}{cc}
1\\
1\end{array}\right)+r_1\left(\begin{array}{cc}
1\\
-1\end{array}\right)=a\left(\begin{array}{cc}
\lambda_1\\
z^+\end{array}\right)+b\left(\begin{array}{cc}
\lambda_1\\
z^-\end{array}\right),
\end{eqnarray}
while at $x=L_1$
\begin{eqnarray}
a\left(\begin{array}{cc}
\lambda_1\\
z^+\end{array}\right)e^{ik_xL_1}+b\left(\begin{array}{cc}
\lambda_1\\
z^-\end{array}\right)e^{-ik_xL_1}=t_1\left(\begin{array}{cc}
1\\
1\end{array}\right).
\end{eqnarray}
Note that the phases from the infinite longitudinal wave vector in the contact regions (where $U_{L,R}\rightarrow\infty$) can be absorbed in the amplitudes $r_1$ and $t_1$. A similar set of equations can be written for a particle incoming from the right contact and transmitted to the left. The resulting scattering matrix has elements
\begin{eqnarray}
t_1&=&t_1'=\frac{\lambda_1 k_x L_1}{D},\\
r_1&=&-r_1'=\frac{-\lambda_1 q L_1 \sin(k_x L_1)}{D},
\end{eqnarray}
where $D\equiv\lambda_1 k_x L_1\cos(k_x L_1)-ik_F L_1\sin(k_x L_1)$, and $ k_x=\sqrt{(\epsilon-u_1)^2/L_1^2-q^2}$.
The index $\lambda_1$ is $+1$ when the Fermi energy is above the onsite potential, or $-1$ when it is below.
The scattering matrix for the second barrier has the same form, with subscript 1 changed to 2. The scattering matrix for the complete double barrier can then be calculated using the composition rule
\[
S^{2B}=
\left(\begin{array}{cc}
r_1+\frac{r_2t_1t'_1}{1-r_2r'_1} & \frac{t'_1t'_2}{1-r_2r'_1}\\
\frac{t_1t_2}{1-r_2r'_1}& r'_2+\frac{r'_1t_2t'_2}{1-r_2r'_1}
\end{array}\right).
\]
We introduce the resulting $S^{2B}$ into Eq.\ (\ref{Qb}), where $\xi_{1,2}=u_{1,2}$, and then take the weak-driving limit of Eq.\ (\ref{weakdriving}).
The index $\lambda_1=\lambda_2$ then only depends on the position of the Fermi energy relative to the working point, and the longitudinal momentum in the pumping regions takes the form $k_x=\sqrt{k_F^2-q^2}$, where $k_F=\epsilon/L=E/(E_L L)$ is the Fermi wave vector.
Collecting all results for the graphene pump with heavily doped contacts (and $L_1=L_2=L$), Eq.\ (\ref{weakdriving}) yields
\begin{eqnarray}\label{QGE}
   Q^{\mathrm{gr-\infty}}_q&=&\pm e A_u\frac{k_FL}{\pi}\frac{(qL)^2}{k_xL}\\
   &\times&\frac{\sin^2(k_xL) \left[\sin (2k_xL)-2k_xL\cos (2k_xL)\right]}{\left[(k_xL)^2+(qL)^2\sin^2 (2k_xL)\right]^2},\nonumber
\end{eqnarray}
where the $\pm$ sign denotes whether the pump is doped with electrons ($+$)
or holes ($-$). This result not only applies to propagating modes (real momentum $k_x$,  $|q|<k_F$),
but also to evanescent modes, $|q|>k_F$, for which $k_x$ is imaginary.

In contrast, a weakly driven graphene pump with undoped leads has
no incoming lead modes that become evanescent in the pump. This is because the Fermi momentum in the pump at the
working point is identical to the Fermi momentum in the contacts ($U_{L,R}=0$). Taking the appropriate limit of the wave-matching results,
each propagating mode then contributes a pumped charge
\begin{equation} Q^{\mathrm{gr-0}}_q=eA_u\frac{k_FL}{\pi}\frac{2(qL)^2\cos(k_xL)\sin^3(k_xL)}{(k_xL)^4},
\label{QGZ}
\end{equation}
where $|q|\leq k_F$ such that $k_x$ is real; there is no contribution by modes with $|q|> k_F$. 

For normal pumps, the wave matching procedure is modified due to the different dispersion relations and absence of the pseudospin degree of freedom;
otherwise, the formalism remains unchanged. In terms of the longitudinal momentum $K_x$ of electrons in the leads, the pumped charge in each channel then takes the form
\begin{eqnarray} \label{eq:normalpumpsq}
	Q^{\mathrm{n}}_q&=&eA_u 8\frac{k_F^2K_x^3}{k_xL^2}\sin^2(k_xL)\\
	&\times& \frac{2k_xL(k_x^2-K_x^2)\cos(2k_xL)+(k_x^2+K_x^2)\sin(2k_xL)}{\left[4 k_x^2K_x^2\cos^2(2k_xL)+(k_x^2+K_x^2)^2\sin^2(2k_xL)\right]^2}
.
\nonumber
\end{eqnarray}
Both the heavily doped contact ($Q^{\mathrm{n-\infty}}_q$) and undoped
contact  ($Q^{\mathrm{n-0}}_q$) normal cases follow from the above expression by
taking  $K_x$ to infinity or $k_x$, respectively.

Equipped with Eqs.\ (\ref{QGE}), (\ref{QGZ}), and the two appropriate limits of Eq.\ (\ref{eq:normalpumpsq}),
we now can compare the results for the different settings.
In all four cases, the pumped charge has a prefactor $A_u k_F$,
indicating that pumping is proportional to the dimensionless driving strength $A_u$ and the pump's number
$N_p=g k_F W/\pi$ of propagating modes at the Fermi energy. The latter number includes
a degeneracy factor $g=4$ for graphene (accounting for two valleys and two physical spin states),
and $g=2$ for normal conductors (spin). By factoring out these
two quantities, we obtain the dimensionless pumping response
\begin{equation}\label{chi}
  \chi_q^u\equiv \frac{\partial_{u}^2 Q_q}{e N_p}\approx\frac{Q_q}{e A_u
  N_p},
\end{equation}
which depends only on the system's scattering characteristics at a
given energy.

\begin{figure}[top]
\includegraphics[width=\linewidth]{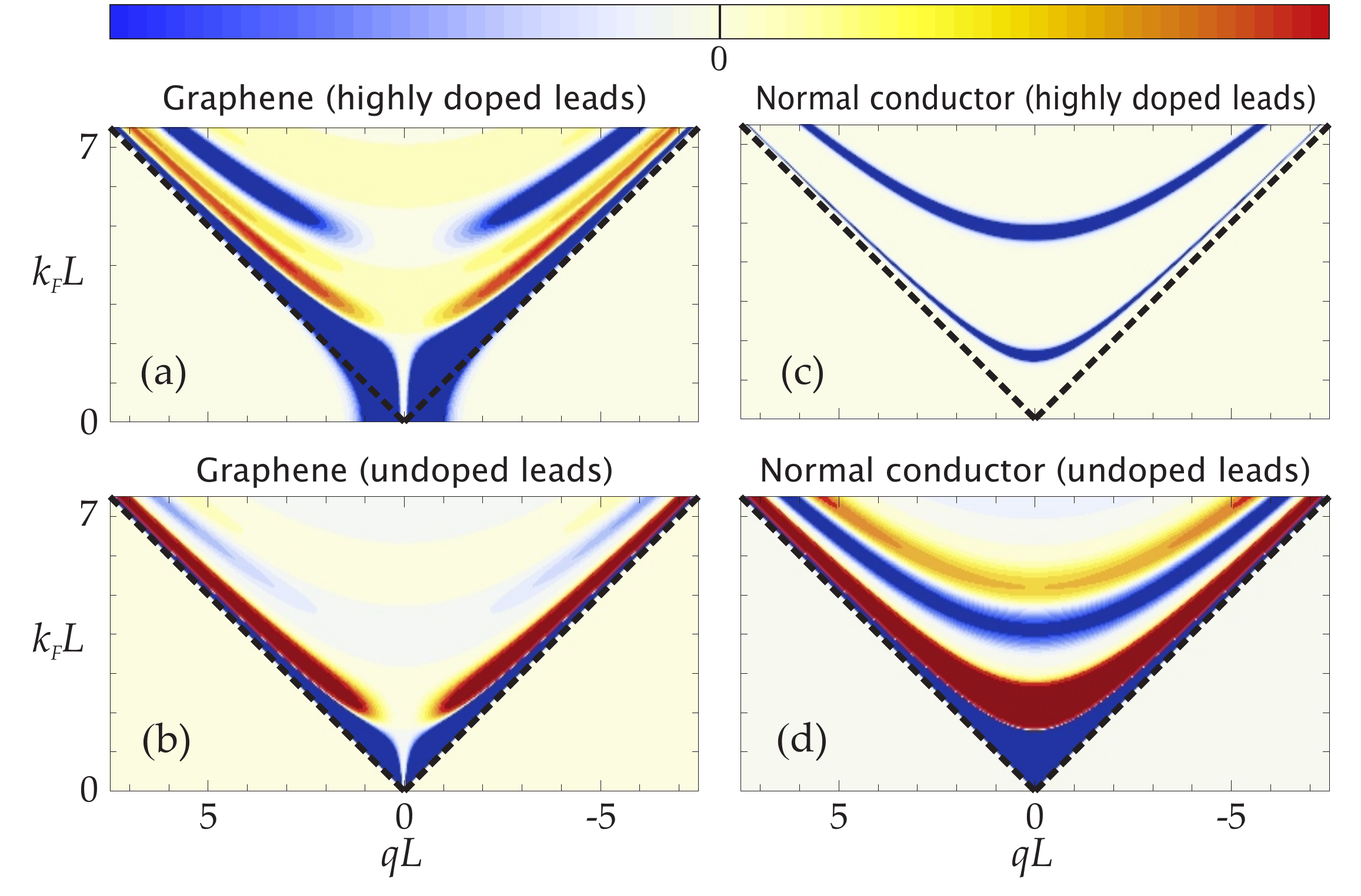}
\caption{\label{fig:chiq}(Color online) Momentum distribution of
pumped charge per mode $\chi^u_q=Q_q/(eA_uN_p)$ as a function of mode index $q$
for varying carrier concentration (parameterized by the Fermi momentum $k_F$).
Blue and red represent opposite directions of pumping (left to
right or right to left).  In graphene, the propagating mode with
$q=0$ (normal incidence) cannot be pumped due to the Klein
paradox. In the case of graphene with heavily doped leads significant pumping
is possible due to the contribution of the evanescent modes
($|q|>k_F$, delineated by the dashed line), which dominate around
the Dirac point ($k_F=0$). The other pumps can only drive current
through the propagating modes.}
\end{figure}

For short and wide systems ($W\gg L$), the dimensionless pumping response develops a quasi-continuous dependence on the transverse momentum $q$.
In the two-dimensional contour plots of Fig.\ \ref{fig:chiq} we examine this dependence for varying carrier concentration (parameterized by the Fermi momentum $k_F$),
comparing the results for the two graphene setups (panels a and b) to those of the two normal pumps (panels c and d).
For graphene, we only show the results for a Fermi energy above the working point;
when the carriers in the central pump region are changed
from electrons to holes, the pumped
current reverts sign if the leads are heavily doped  [cf.\ Eq.\ (\ref{QGE})], but remains the same if the
leads are undoped [in any case the pumped current always changes sign when one
reverts the pumping cycle].

In each panel, the dashed line $|q|=k_F$ delineates the border between propagating modes ($|q|<k_F$) and
evanescent modes ($|q|>k_F$). Pumps with undoped contacts cannot access such modes.
For highly doped contacts, evanescent modes penetrate into the pumping region,
but only in the case of graphene  [Fig.\ \ref{fig:chiq}(a)] their contribution is sizeable. This is especially true
close to the Dirac point, where the evanescent modes dominate.

This effect can be understood by considering the
specific  conditions for electronic confinement in graphene, which are directly linked to the chirality of the charge carriers.
Chirality conservation at the contact enables evanescent electrons to
populate the graphene pumping region for modes within a window of width
$\Delta q\sim 1/2L$ around $q=0$ \cite{Katsnelson:EPJB06,Tworzydlo:PRL06,Schomerus:PRB07}. These evanescent modes contribute to pumping
because they are sensitive to the onsite potentials $U_i$ and have a finite
amplitude at both contacts, so that charge transfer between them is possible
over a pumping cycle. It is noteworthy that, in contrast, the propagating mode
with $q=0$ (normal incidence) never contributes to the pumping in graphene [both for doped as well as for undoped contacts, Fig.\ \ref{fig:chiq}(b)].
This is a direct consequence of Klein tunneling \cite{Katsnelson:NP06}---the transmission
$T_{q=0}=1$ is perfect at all energies, the mode is therefore insensitive to driving and cannot be pumped.

Carriers in normal conductors do not display chirality, which entails that they
can be easily confined. In particular, for a normal conductor with heavily doped contacts, the
large Fermi velocity mismatch suppresses the transparency of the contacts for
all modes except those close to resonance with well-resolved energy levels.
In Fig.\ \ref{fig:chiq}(c), these levels are seen as narrow regions of finite pumping,
along with a threshold $k_F L= \pi/2$ below which no pumping occurs.
The pumping is
directed, meaning that for a given orientation of the driving cycle, the
pumped current has the same sign for all energies.
The different confinement
properties also entail that the contribution of evanescent modes
to pumping is negligible at all energies.

The normal pumps become open when they are attached to undoped leads, and
consequently in this case there is no energy threshold for pumping
[Fig.\ \ref{fig:chiq}(d)]. The sign of the pumped
current is energy dependent, which is a generic feature of open pumps
(including the graphene pump with doped leads). However, the contribution of
evanescent modes vanishes identically since all incoming modes remain
propagating in the pumping region.

\begin{figure}[top]
\includegraphics[width=0.9\linewidth]{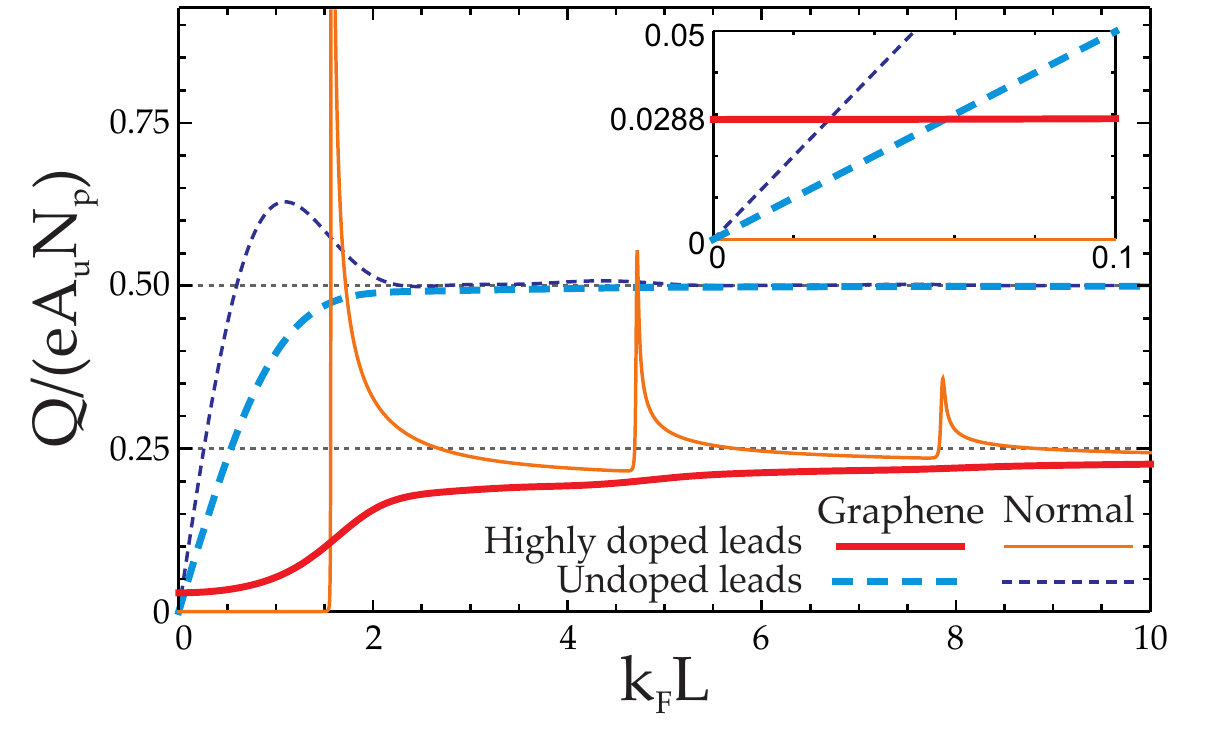}
\caption{\label{fig:chi} (Color online) Pumped charge $Q$ per cycle (normalized to the pumping strength $A_u$ and the number of propagating modes
$N_p$ in short and wide pumps, $W\gg L$),
as a function of the carrier concentration (parameterized by the Fermi momentum $k_F$).
The plotted quantity is the pumping response $\chi^u$, introduced in Eq.\ (\ref{chisum}),
which is evaluated  using Eq.\ (\ref{QGE}) (graphene with highly doped leads, thick solid curve),
Eq.\ (\ref{QGZ}) (graphene with undoped leads, thick dashed curve),
and the appropriate limits of Eq.\ (\ref{eq:normalpumpsq}) (normal pump with highly doped leads, thin solid curve, and
with undoped leads, thin dashed curve). The inset focusses onto the region close to charge neutrality ($k_FL\ll 1$),
where evanescent modes in the
graphene pump with highly doped leads allow for a finite charge transfer  approaching the
universal value of $\chi^u=0.0288$.}
\end{figure}

The total pumped charge can be characterized by
summing the mode-resolved result over all incoming modes,
\begin{equation}\label{chisum}
  \chi^u=g\sum_q \chi_q^u=\frac{g}{eN_p}\sum_q \partial_{u}^2 Q_q\approx \frac{Q}{e A_u N_p}.
\end{equation}
For short and wide systems ($W\gg L$), the sum can be approximated by an
integral over the continuous transverse momentum, $\sum_q \to (W/2\pi)\int dq$.

The result for the four types of pumps is shown in Fig.\ \ref{fig:chi}.
For large energies ($k_F\gtrsim 1/L$) the
pumping response rises to $1/2$ and $1/4$ in the cases of undoped and highly
doped leads, respectively. As a consequence of the contact-induced resonant tunneling subbands,
the normal pump with highly doped leads
only operates above a finite carrier-concentration threshold.
As highlighted in the inset, the evanescent electron pumping in graphene with highly doped leads
results in a finite pumped charge at nominally vanishing charge-carrier density ($k_F=0$).
In the considered limit $W\gg L$, the response is independent of the detailed system characteristics, like the system dimensions, aspect ratio (as long as it is large), or the precise position of the Fermi energy (as long as it is within less than $E_L$ of the pump's Dirac point). It then acquires the universal dimensionless value
\begin{equation}
  \int_0^\infty dq \frac{\sinh^2(q)\left[2q\cosh(2q)-\sinh(2q)\right]}{\pi q^3 \cosh^4(2q)}
  =0.0288.
\end{equation}
Due to its association to evanescent modes, this value can be interpreted as the pumping analogue to the minimal conductivity in stationary transport
\cite{Katsnelson:EPJB06,Tworzydlo:PRL06,Schomerus:PRB07}. All other pumps have a
vanishing pumping response at $k_F=0$.

\section{\label{sec:4}Gate voltage driving}

In a realistic experimental pumping setup, the
principal driving parameters are not the onsite energies $U_i$, but gate
voltages $V_i$ (see Fig. \ref{fig:system}) which control the locally induced charge densities $\rho_i$.
The  onsite energy is related to the charge density via the density of states, which differs between normal systems and graphene.
In particular, the compressibility of graphene vanishes at the Dirac point \cite{Martin:NP08}.
In the following we will see that this enhances the response of a graphene pump.

Because of screening, the translation between $V$ and $U$ in general requires a self-consistent treatment since the equilibrium charge density profile, and hence $U(x)$, can be a complicated non-local function of $V(x)$.  The charge in the pump, which interacts with the whole gate structure, equilibrates to the density $\rho(x)$ that minimizes the total electrostatic energy. As a consequence, the onsite potential profile created by spatially piecewise constant gate voltages will acquire deviations from the piecewise constant model we have assumed for $U(x)$ in the previous sections. A second non-local screening contribution comes from the influence of the lead electrons on the density close to the contact regions. This `lead-doping' effect is particularly relevant in graphene, which has more transparent contacts than a normal pump. As a result, the equilibrium electron density profile varies smoothly across the transparent contact from the lead's high density to the lower density in the pump, which results in an inhomogeneous charge density (and hence onsite potential) close to the contact. The equilibrium distribution depends on screening and contact details and was studied within an ab-initio approach in Ref.\ \cite{Barraza-Lopez:PRL10}. The screening of $U(x)$ is expected to modify the results of the abrupt barrier model only quantitatively, since perfect normal transmission is preserved, although the angular profile away from normal incidence is narrowed.

In view of these complications, we rely on the large capacitance of the metallic gates to ignore the detailed local effects of screening in the pumping region, and account for non-local screening effects by relating the total charge $n_i$ under gate $i=1,2$ to voltages $V_j$ through a non-diagonal capacitance matrix $C_{ij}$, $n_i(t)= \sum_j C_{ij}V_j(t)$ (zero voltage is identified to charge neutrality).
We then are in a position to assume that the charge density $\rho_i$ under each gate electrode is constant.
The local onsite energy is related to the charge density by the integral of the density of states from the local position of the neutrality point to the Fermi energy.
Taking into account the different densities of states of graphene and normal conductors, one can then express the total charge $n_i$ in terms of the dimensionless onsite potential $u_i\equiv U_i/E_L$ and Fermi energy $\epsilon\equiv E/E_L$ used in our previous analysis. For graphene $n_i(t)=e W (\epsilon-u_i(t))^2/(\pi L_i)$, whilst for the normal case $n_i(t)=e W (\epsilon-u_i(t))/(2\pi L_i)$. We then have
\begin{eqnarray}
u_i-\epsilon=
\left\{\begin{array}{ll}
-\sqrt{\frac{L}{W}\frac{\pi \sum_j C_{ij}V_j}{e}}&\mbox{graphene}\\
-\frac{L}{W}\frac{2\pi \sum_j C_{ij}V_j}{e}&\mbox{normal}
\end{array}
\right.
\end{eqnarray}

If we drive the pump with a weak adiabatic gate voltage cycle $V_i(t)=V_i(0)+\delta V_i(t)$, with $V_i(0)$ such that the working point is uniform, $u_1(0)=u_2(0)=0$ [and therefore $n_1(0)=n_2(0)=n(0)$], we can apply the theory of the previous sections with a corresponding cycle in the onsite potential parameter space $u_i(t)=(d u_i/dV_j)\delta V_j(t)$.
The resulting areas enclosed in the $u$ and $V$ parameter spaces are then related by the Jacobian $A_u=A_V\det\left(\partial u_i/\partial V_j\right)$. Specifically
\begin{equation}
A^{
\mathrm{gr}}_u=\frac{\pi}{4} \frac{L}{W}\frac{1}{e n(0)}\det (C_{ij})A_V
\end{equation}
in graphene, and
\begin{equation}
A^{
\mathrm{n}}_u=4\pi^2\frac{L^2}{W^2}\frac{1}{e^2} \det (C_{ij})A_V
\end{equation}
for the normal pump. Note the $n(0)$ in the denominator of $A^\mathrm{gr}$. This implies that the onsite energy response to external gate driving in graphene diverges at the neutrality point, and can be seen as a direct consequence of the vanishing electronic compressibility, $\kappa\propto(\partial u/\partial V)^{-1}\rightarrow 0$.
Furthermore, we observe that the pumped current in the weak driving regime $A_u\ll 1$ is suppressed by the cross capacitance between the electrodes (since they diminish $\det C_{ij}$). The optimal driving is still achieved by maximizing the area $A_V$, which for the typical case of sinusoidal driving corresponds to a phase shift of $\pi/2$ between the two gates.

The pumped current at driving frequency $\nu$ is given by $I=\nu Q=e\nu A_u N_p\chi^u(k_F L)$. We can express $I$ as a function of $A_V$ and the initial charge under each gate, $n(0)=\sum_j C_{ij}V_j(0)$, by using
\begin{equation}
N_p=\frac{g}{\pi} k_F W=2 \sqrt{\frac{g}{\pi}  \frac{W}{L}\frac{n(0)}{e}},\quad
k_F L=2\sqrt{\frac{\pi}{g} \frac{L}{W}\frac{n(0)}{e}}.
\end{equation}
This gives for graphene
\begin{equation}
I^\mathrm{gr}=e\nu\times\sqrt{\pi}\left(\frac{L}{W}\right)^{1/2}
\frac{A_V\det C_{ij}}{e^2}\sqrt{\frac{e}{n(0)}}\chi^\mathrm{u},
\end{equation}
whilst for a normal pump
\begin{equation}
I^\mathrm{n}=e\nu \times 8\sqrt{2}\pi^{3/2}\left(\frac{L}{W}\right)^{3/2}
\frac{A_V\det C_{ij}}{e^2}\sqrt{\frac{n(0)}{e}}\chi^\mathrm{u}.
\end{equation}
When driving via gate voltages at fixed strength $A_V$, the pumping efficiency $I^\mathrm{gr}/A_V$ diverges in graphene close to its incompressible neutrality point ($n(0)\rightarrow 0$), while it vanishes in the normal case. Recall, however, that by assumption $\delta V_i(t)\ll V_i(0)$, so that $A_V\det C_{ij}\ll n^2(0)$. Therefore, although the pumping efficiency for weak driving may diverge in graphene, the current itself will not.

\section{\label{sec:5}Pumping limitations}

As a general rule, the pumped current increases with increasing pumping frequency $\nu$ for small $\nu$. In the adiabatic regime studied here it does so linearly in $\nu$, but this optimal scaling typically becomes  sublinear as one approaches the non-adiabatic regime, $h\nu\sim E_L$, \cite{San-Jose:11} possibly with superimposed oscillations depending on the model \cite{Buttiker:QMath06} . For a small graphene pu  mp of $L=0.1\mu$m, this adiabatic frequency ceiling is around $\nu_{0.1\mu\mathrm{m}}\sim 1.6$ THz, which drops down to $\nu_{1\mu\mathrm{m}}\sim 0.16$ THz for a larger $L=1\mu$m pump. This is a very high frequency if we compare it to the one of normal pumps, $\nu_{0.1\mu\mathrm{m}}=14$ GHz and $\nu_{1\mu\mathrm{m}}=0.14$GHz.

Regarding the maximal magnitude of adiabatically pumped current, both normal (fabricated e.g. on GaAs/AlGaAs heterojunctions) and graphene pumps are comparable, providing an estimated 10 to 500 nA for typical setups with many electrons in the pump. Assuming $W/L=6$, $C\sim 1$ aF and $V(0)\sim 10$V (which corresponds to an $n(0)$ of around 62 electrons under each gate) and a driving of $\delta V\sim 1$V, we have a maximum current of around 225 nA (725 nA) for graphene (normal) pumps of length $L=0.1\mu$m, which drops down to 23 nA (7 nA) for $L=1 \mu$m. The advantage of graphene in this respect becomes most noticeable in the few-electron regime $n(0)\sim 1$. As an example, the maximum current in an $L=0.1 \mu$m graphene pump at $V(0)\sim 10~ \delta V\sim 1$ V is still in the $0.8$ nA range due to the evanescent-mode contribution, while the normal pump is effectively inoperative in this range, due to the first subband threshold. A further  advantage of graphene versus normal pumps is that the ballistic transport regime faborable for pumping which is assumed throughout this work is much more easy to achieve (especially at high temperatures) in graphene than in semiconducting heterostructures.

We finally discuss thermal fluctuations, which generally degrade quantum pumping (possible exceptions are Refs. \cite{Moskalets:PRL08, Moskalets:PRB08} ). At the very least, if $T<h\nu/k_B$, one would just observe an effective thermal smearing of $\chi(k_F L)$ in Fig.\ \ref{fig:chi}, since in this regime thermal fluctuations can be considered static within each driving cycle \cite{Vavilov:PRB01}. Above this temperature,  thermal fluctuations will quickly suppress quantum pumping, since quantum coherence within a single pumping cycle is destroyed. Since $\nu$ has a ceiling given by adiabaticity, this allows us to estimate the maximum operating temperature of adiabatic pumps by $T_L=E_L/k_B$, which is around 76 K for a small $L=0.1 \mu$m graphene pump, as opposed to $7.6$ K for a  comparable normal pump. This thermal advantage arises as a result of the density of states. It is also one of the reasons for the temperature robustness of other transport effects in graphene.

\section{\label{sec:6} Concluding remarks}

In summary, we have investigated fundamental and practical aspects of adiabatic quantum pumping which distinguish graphene-based systems
from equivalent setups involving conventional two-dimensional electron gases.
Because of the unique  properties of graphene (in particular, the chirality of the charge carriers), 
evanescent modes can contribute significantly to the pumping, especially when the system
is operated close to the charge-neutrality point. For the case of short
and wide pumps, the evanescent pumping regime is characterized by a universal value of the dimensionless pumping response. This value does not depend on the width or length of the pump. It is also largely independent of temperature, as long as $T<E_L/k_B$, since the evanescent mode pumping response is quite flat within energies $\sim E_L$ of the Dirac point.
In normal pumps evanescent modes only give a negligible contribution.

In practical terms, the vanishing electronic compressibility of graphene at the Dirac point enhances the 
response of graphene-based pumps to driving via external gate potentials.
This presents a clear performance advantage for graphene pumps if they are operated in the few electron regime. 
Furthermore, graphene-based quantum pumping promises an enhanced robustness against thermal effects, as already known from stationary transport. 
For the same reason, the regime of adiabatic driving extends to higher frequencies than in normal pumps.
As for other graphene-based electronic applications, these attractive features are further enhanced by the long coherence time and high mobility 
of charge carriers in graphene.

We gratefully acknowledge financial support from MICINN (Spain), through grants FIS2009-08744 and FIS2008-00124, and the support from the European Commission, Marie Curie Excellence Grant MEXT-CT-2005-023778.

\bibliography{biblio}

\end{document}